\documentclass[letterpaper]{jpconf}

\usepackage{citesort} 

\begin{document}

\title{\bfseries Constrained dynamics of universally coupled massive spin 2-spin 0 gravities }

\author{  J. Brian Pitts}
\address{ History and Philosophy of Science Graduate Program, University of Notre Dame, 
Notre Dame, Indiana 46556  USA   }
\ead{jpitts@nd.edu}


\begin{abstract} The 2-parameter family of massive variants of Einstein's gravity (on a Minkowski background) found by  Ogievetsky and Polubarinov by excluding lower spins can also be derived using universal coupling.  A Dirac-Bergmann constrained dynamics analysis seems not to have been presented for these theories, the Freund-Maheshwari-Schonberg special case, or any other massive gravity beyond the linear level treated by Marzban, Whiting and van Dam.    Here the Dirac-Bergmann apparatus is applied to these theories. A few remarks are made on the question of positive energy.  Being bimetric, massive gravities have a causality puzzle, but it appears soluble by the introduction and judicious use of gauge freedom. 
\end{abstract}


\section{Universal Coupling and Massive Gravity: Historical Sketch}

Although the field approach to Einstein's equations is sometimes contrasted with the  geometrical  approach to gravitation (supposedly Einstein's), in fact core ideas of the field approach, such as the gravitation-electromagnetism analogy, universal coupling to a combined matter-gravity energy-momentum complex as source, and conservation (in the sense of an ordinary divergence) of energy-momentum due to the gravitational field equations alone, were employed by Einstein in the 1910s \cite{EinsteinPapers4,EinsteinPapers8,Janssen,PrincipleRelativity} in his quest for the gravitational field equations.  The field approach enjoyed a revival in the 1950s-70s \cite{NullCones}, especially in works by   Kraichnan \cite{Kraichnan},  Gupta \cite{Gupta},  Feynman \cite{Feynman}, and  Deser \cite{Deser}. In the 1960s V. I. Ogievetsky and I. V. Polubarinov (OP) derived a 2-parameter family of Poincar\'{e}-symmetric massive Einstein's equations. Universal coupling was not used, but the spin 1 degrees of freedom were removed from the interacting theory to avoid negative energies and one spin 0 to preserve locality \cite{OP}. Thus spin 2 and one spin 0 degrees of freedom remained. Independently,   Freund,   Maheshwari and  Schonberg (FMS) derived perhaps the most attractive member of the OP family using universal coupling with a mass term included \cite{FMS}.  Previously the author and W. C. Schieve showed that two one-parameter subfamilies of OP theories are universally coupled \cite{PittsMassive}. Recently (unpublished) the author showed that all OP theories are universally coupled.  Empirically, massive spin 2-spin 0 gravity matches  GR except for strong fields or large distances \cite{Visser,GrishchukMass}. The spin 0, which is repulsive, theoretically can have a mass anywhere between 0 and $\infty,$ including both endpoints, so its phenomenology has some flexibility \cite{GrishchukMass}.

Two further questions arise: positive energy and causality.  In the late 1930s Pauli and  Fierz, 
 working to linear order, argued that massive spin 2 theories ought to have no spin 0, because the latter was of the wrong sign and so threatened positive energy.  While OP seem to have ignored this problem for their nonlinear theories, FMS suggested that subtle effects might render the wrong-sign spin 0 harmless.  In 1970 the van Dam-Veltman-Zakharov discontinuity was derived: at linear level, massive spin 2 gravity with no spin 0 disagrees in the massless limit with GTR and, more seriously, with experiment \cite{vDVmass2,Zakharov}.  Boulware and  Deser argued \cite{DeserMass} that massive gravity was a doomed enterprise because every theory suffered from either violation of positive energy (the spin 2-spin 0 case) or  empirical refutation (the spin 2 case), or both.  This argument was widely but not quite universally accepted for over two decades.  Since the mid-1990s both the empirical inadequacy of spin 2 theories \cite{Vainshtein2} and the negative energy instability of spin 2-spin 0 theories \cite{Visser,GrishchukMass} have been discussed anew. The former question is of little relevance here (except perhaps to the infinitely massive spin 0 case).  About the latter question I have little to say  here, though elsewhere  \cite{PittsMassive} I have suggested  that nonperturbative features of the Hamiltonian should not be ignored and might be of some help. In any case, the spin 2-spin 0 theories discussed here are viable only if the negative energy worry can be handled.  While no one has proven that these theories are stable, recently Visser and Babak and Grishchuk have suggested that they might be \cite{Visser,GrishchukMass}.  The issue of causality, which was not considered until the last two decades, will be discussed at the end.

\section{Ogievetsky-Polubarinov  Massive Gravities}

Some mass terms are better motivated than others.  The OP mass terms are well motivated.  The author's recent rederivation of the OP massive gravities using universal coupling  (extending that in  \cite{PittsMassive} is based on the metrical rather than canonical stress tensor, along the lines of Kraichnan \cite{Kraichnan} and Deser \cite{Deser} rather than Gupta \cite{Gupta} and FMS \cite{FMS}, so formal general covariance is achieved with a flat metric tensor (under arbitrary coordinate transformations) $\eta_{\mu\nu}.$ (By contrast, OP used Cartesian coordinates with imaginary time.) The OP theories are best expressed using tensor densities of arbitrary real weight.  The flat metric's weight $-l$ covariant concomitant is $ \tilde{ \eta}_{\mu\nu} = \sqrt{-\eta} \: ^{-l} \eta_{\mu\nu}$; its inverse,  the  weight $l$ contravariant concomitant, is $ \tilde{ \eta}^{\mu\nu} = \sqrt{-\eta} \: ^{l} \eta^{\mu\nu}$.  The densitized curved metric $\tilde{g}_{\mu\nu}$ and its inverse $\tilde{g}^{\mu\nu}$ are analogously defined in terms of $g_{\mu\nu}.$  In addition to the density weight $l,$ the OP theories are parametrized by another parameter $n$ ($l \neq \frac{1}{2}, n \neq 0$).  The two metrics are connected by the equation $\tilde{g}^{\mu\nu} = \tilde{\eta}^{\mu\nu} + \lambda \tilde{\gamma}^{\mu\nu},$ where  $\tilde{\gamma}^{\mu\nu}$ is the (weight $l$) gravitational  potential  and $\lambda= -\sqrt{32\pi G}.$  The parameter $n$ gives the power (in the sense of a binomial series) to which 
$\tilde{g}^{\mu\nu}$ is raised:  $$ [ \tilde{g}^{\mu\nu}]^{n} = \tilde{\eta}^{\mu\nu}  + n \lambda  \tilde{\gamma}^{\mu\nu} +  \frac{n(n-1)}{2!} \lambda^{2} \tilde{\gamma}^{\mu\alpha} \tilde{\eta}_{\alpha\rho} \tilde{\gamma}^{\rho\nu} + \ldots  
$$  Thus, for example, $[ \tilde{g}^{\mu\nu}]^{2} =  \tilde{g}^{\mu\alpha} \tilde{\eta}_{\alpha\rho} \tilde{g}^{\rho\nu}$ and $[ \tilde{g}^{\mu\nu}]^{-1} =  \tilde{\eta}^{\mu\alpha} \tilde{g}_{\alpha\rho} \tilde{\eta}^{\rho\nu} $.  In case this series diverges, these arbitrary real powers can be defined using a  generalized eigenvalue formalism  based on the Segr\'{e} classification of $g_{\mu\nu}$  with respect to $\eta_{\mu\nu}$  \cite{NullCones1}.

The action for the $S$ for the OP theories, once stripped of parts that contribute nothing to the equations of motion, is 
$$ \frac{1}{16 \pi G} \int d^{4}x \sqrt{-g} R(g)  + S_{matter}[g_{\mu\nu}, u] 
 + \frac{m^2}{32 \pi G n} \int d^{4}x ( 2\sqrt{-g}[2l-1]  - \frac{1}{n}\sqrt{ -\eta} [\tilde{g}^{\mu\nu}]^{n} \tilde{\eta}_{\mu\nu} 
).$$  The mass term contains a formal cosmological constant term $\sqrt{-g}$ and the all-important new piece 
involving $- \frac{1}{n}\sqrt{ -\eta} [\tilde{g}^{\mu\nu}]^{n} \tilde{\eta}_{\mu\nu},$  which ensures Yukawa exponential fall-off (rather than the increase with distance that a  cosmological constant produces), makes the flat metric observable in principle (though empirically the graviton mass is tiny if nonzero),  and destroys the gauge freedom.  The terms in the action which do not contribute to the field equations include a constant term that ensures vanishing energy for Minkowski space-time and a divergence term.  The case   $l=1, n=1$ gives the FMS theory, which has also been adopted by  Logunov \cite{LogunovBook}.


\section{Constrained Dynamics of OP Massive Gravities} 

While some  Hamiltonian treatment of massive GR was given by Boulware and Deser \cite{DeserMass} and the Dirac-Bergmann constrained dynamics formalism was applied to linearized massive GR by Marzban, Whiting, and van Dam \cite{Marzban}, apparently no explicit constrained dynamics treatment of nonlinear massive variants of Einstein's equations has been given. Although Marzban, Whiting and van Dam found the linear spin 2 theory to have some nontrivial features in the constrained dynamics analysis, the linear   spin 2-spin 0 theories  behave much as one would expect from the Proca massive electromagnetic analog with mass term $-\frac{1}{2}m^2 A^{\mu} A_{\mu}$. It is useful to see what differences an exact (nonlinear) treatment of gravity gives.  I follow the standard GTR treatment \cite{Sundermeyer}. Some terms that do not affect the field equations---constants and divergences---will be dropped from the Lagrangian density and an ADM split will be employed.  For isolated sources without radiation, one expects exponential Yukawa falloff, so discarding boundary terms for massive theories is less dangerous than it is for GR. The notation of lapse $N$, shift vector $\beta^{i}$, curved spatial metric $h_{ij},$ and extrinsic curvature $K_{ij}$ is employed, along with the pseudo-vectorial abbreviation  $N^{\mu}= (N, \beta^{i}).$   I set  $16 \pi G = 1.$  For the flat metric, Cartesian coordinates are chosen so $\eta_{\mu\nu}=diag(-1,1,1,1).$ The symbol $\approx$ means  equality on the constraint surface.

The Lagrangian density from which the Hamiltonian density will be found is therefore
   \begin{eqnarray*}     \mathcal{L} = N\sqrt{h}  (K^{ij} K_{ij} -K^{i}_{i}  K^{j}_{j} + R[h_{ij}]) 
+ \frac{m^2}{32 \pi G n}  ( 2\sqrt{-g}[2l-1]  - \frac{1}{n} [\tilde{g}^{\mu\nu}]^{n} \tilde{\eta}_{\mu\nu} 
).    \end{eqnarray*} The conjugate momenta are the same as in GR:  $\Pi^{ij} = \frac{  \partial \mathcal{L} }{  \partial \dot{h}_{ij}  };$ $P_{\mu}(x) =  \frac{  \partial \mathcal{L} }{  \partial \dot{N}^{\mu} } = 0$ are primary constraints.
The generalized Legendre transformation gives the canonical Hamiltonian density  \begin{eqnarray*} 
\mathcal{H} = \Pi^{ij} \dot{h}_{ij}  +  P_{\mu} \dot{N^{\mu}} - \mathcal{L} 
 = \Pi^{ij} \dot{h}_{ij}  +   - \mathcal{L} 
=N \mathcal{H}_{0} + \beta^{i} \mathcal{H}_{i}   
- \frac{m^2}{32 \pi G n}  ( 2\sqrt{-g}[2l-1]  - \frac{1}{n} [\tilde{g}^{\mu\nu}]^{n} \tilde{\eta}_{\mu\nu} 
),   \end{eqnarray*}
where $\mathcal{H}_{0},$ $ \mathcal{H}_{i}$ are the familiar functions of $h_{ij},$  $\Pi^{ij},$  and their  spatial derivatives.  Using $\sqrt{-g} = N \sqrt{h}$, one can absorb $ - \frac{ [2l-1]  m^2}{16 \pi G n}  \sqrt{-g} $ into $ \tilde{ \mathcal{H}}_{0}$ without affecting the Dirac `algebra'  of Poisson brackets 
\cite{Sundermeyer}. Having done so, I now write  $\mathcal{H}_{ml}$ for the leftover piece of the mass term, $ \frac{m^2}{32 \pi G n^2}   [\tilde{g}^{\mu\nu}]^{n} \tilde{\eta}_{\mu\nu}.$   Thus $\mathcal{H} = N \tilde{\mathcal{H}_{0}} + \beta^{i} \mathcal{H}_{i}   
+ \mathcal{H}_{ml}.$  As usual demanding the preservation of the primary constraints  $P_{\mu}=0$  gives secondary constraints
$ \frac{ \partial \mathcal{H} }{\partial N }  = \tilde{\mathcal{H}_{0}}   +
		\frac{ \partial \mathcal{H}_{ml}  }{\partial N }, 
 \frac{ \partial \mathcal{H} }{\partial \beta^{i} }  = \mathcal{H}_{i}    +
		\frac{ \partial \mathcal{H}_{ml}  }{\partial \beta^{i} }.$   For $n = \pm 1,$ it is straightforward to treat $\mathcal{H}_{ml}$ explicitly, but not in general.  Preserving  the (smeared) secondary constraints 
  $\int d^{3}x \frac{ \partial \mathcal{H} }{\partial N^{\mu} } \xi^{\mu}(x)$ involves the primary Hamiltonian  $ \int d^{3}y (N \tilde{\mathcal{H}_{0}} + \beta^{i} \mathcal{H}_{i} + \mathcal{H}_{ml}  + U^{\mu} P_{\mu})$
with Lagrange multipliers $U^{\mu}$.     Now one uses the Dirac algebra.
If $  \frac{ \partial^2 \mathcal{H} }{\partial N^{\mu} \partial N^{\nu} },$ the coefficient of  $U^{\mu},$ is an invertible matrix, then no tertiary constraints arise.

  Are constraints all second class?  The answer depends on the determinant of the matrix of Poisson brackets,
\begin{eqnarray*}  
det \left[ \begin{array}{cc} 
\{ P_{\mu}(x), P_{\nu}(y) \} 	 &  \{ P_{\mu}(x), \frac{ \partial \mathcal{H} }{\partial N^{\nu} }(y) \}      \\
\{  \frac{ \partial \mathcal{H} }{\partial N^{\mu}}(x), P_{\nu}(y)   \}   & \{  \frac{ \partial \mathcal{H} }{\partial N^{\mu}}(x),    \frac{ \partial \mathcal{H} }{\partial N^{\nu} }(y) \}   
\end{array} \right]    =   
det \left[ \begin{array}{cc} 
	0 &  \{ P_{\mu}(x), \frac{ \partial \mathcal{H} }{\partial N^{\nu} }(y) \}      \\
\{  \frac{ \partial \mathcal{H} }{\partial N^{\mu}}(x), P_{\nu}(y)   \}   &  \{  \frac{ \partial \mathcal{H} }{\partial N^{\mu}}(x),    \frac{ \partial \mathcal{H} }{\partial N^{\nu} }(y) \}   
\end{array} \right]    =  \\ 
det \left[ \begin{array}{cc} 
	0 &  \{ P_{\mu}(x), \frac{ \partial \mathcal{H} }{\partial N^{\nu} }(y) \}      \\
\{  \frac{ \partial \mathcal{H} }{\partial N^{\mu}}(x), P_{\nu}(y)   \}   & 0 
\end{array} \right] = 
det \left[ \begin{array}{cc} 
	0 &  -\frac{ \partial^2 \mathcal{H} }{\partial N^{\mu} \partial N^{\nu} } \delta(x,y)     \\ \frac{ \partial^2 \mathcal{H} }{\partial N^{\mu} \partial N^{\nu} } \delta(x,y) 
   & 0 
\end{array} \right].
\end{eqnarray*}
Thus the large difficult determinant reduces in effect to a smaller and simpler  one.  All constraints are second class if the matrix $\frac{ \partial^2 \mathcal{H} }{\partial N^{\mu} \partial N^{\nu} } $ is invertible, which was the same condition that would exclude tertiary constraints.  This matrix is algebraic in the field variables and is determined solely by  $\mathcal{H}_{ml}. $ Due to algebraic complexity, it still is not easy calculate the determinant  in question for general $n.$  However, there are two special cases where it becomes straightforward, $n= \pm 1$ (for which the universal coupling derivation has been presented \cite{PittsMassive}).  The case $n=1,$ which corresponds to a contravariant metric density (to the first power) and includes FMS, gives $ det \sim (l-2)(l-3) + l(l-1)N^{2}h^{ij}\delta_{ij} + \beta^{i}\beta^{i}(l-2)(l-1),$ which is nonzero in general (unless perhaps some singular field configurations could give degenerate cases).  The case $n=-1,$ which roughly corresponds to a covariant metric density, yields  $ det \sim (l-1)(l-2) N^{2} + l(l+1)h_{ij}\delta^{ij} + l(l-1) \beta^{i}h_{ij}\beta^{j},$ which also is nonzero usually or always.  For general $n$ in the weak field limit, one gets $ det \sim (n-1)(4l^2- 4l + 4) + 4l^2 -8l + 6 \neq 0,$ which is nonzero for all values of $l$ and $n$ that exclude tachyons.  Thus even the infinite spin 0 mass case, which OP describe as pure spin 2 \cite{OP}, does not behave like the Marzban-Whiting-van Dam spin 2 case, which has tertiary and quarternary constraints, for a total of 10 second class constraints at each point, leaving 5 degrees of freedom (spin 2). The OP theories have 8 second class constraints and thus so have 6 degrees of freedom (spin 2 and spin 0), in agreement with Boulware and Deser's conclusion that any massive variant of the (nonlinear) Einstein equations must have 6 degrees of freedom \cite{DeserMass}, even if the mass term would yield 5 degrees of freedom with the linearized Einstein equations. Thus the algorithm terminates without tertiary constraints, much as it does for Proca's massive electromagnetism.   The lapse   $N$ and shift $\beta^{i}$  are determined by $h_{ij},$ $\pi^{ij}$ and $\eta_{\mu\nu}.$ Thus the Hamiltonian density    $\mathcal{H} = N \tilde{\mathcal{H}_{0}} + \beta^{i} \mathcal{H}_{i}   
+ \mathcal{H}_{ml} + \ldots$ (constant term and total divergence omitted) can be written partly on-shell as 
$\mathcal{H} \approx  \mathcal{H}_{ml} - N\frac{ \partial \mathcal{H}_{ml} }{\partial N}
 - \beta^{i} \frac{ \partial \mathcal{H}_{ml}  }{\partial \beta^{i} } + \ldots,$ which (apart from the discarded divergence that integrates to 0 for Yukawa exponential fall-off) can be expressed as a fairly simple algebraic function of the two metrics without the momenta.  For the FMS theory ($n=1,l=1$), $ \mathcal{H} \approx \frac{m^2}{16 \pi G}(\frac{\sqrt{h}}{N} -1)$ (plus a divergence).   If the claimed negative energy instability exists, numerical relativists might detect it while working in spherical symmetry.  


\section{Causality, Two Metrics and Gauge Freedom}

 Massive gravity is special relativistic  because $\eta_{\mu\nu}$ is observable in principle, being present in the field equations.  The theories's  symmetry group is the  Poincar\'{e} group (or the conformal group for massless spin 0).  Although nongravitational test bodies follow geodesics of $g_{\mu\nu}$, gravity also sees $\eta_{\mu\nu}.$  Thus the propagation of gravitational or matter radiation  is  subject to the usual special relativistic argument that propagation outside the null cone of $\eta_{\mu\nu}$ in one inertial frame entails backwards causation in another, which presumably one wants to avoid.  It follows that either massive gravity is acausal, or the light cone of $g_{\mu\nu}$ is nowhere outside the null cone of $\eta_{\mu\nu}$ (the condition of $\eta$-causality) \cite{NullCones1,Penrose,Zel2}.  (How seriously one takes this issue depends on whether the Minkowski background is taken as part of fundamental physical law or an approximate and contingent description of phenomena in our region of space-time.  In the former case one might hope to quantize using the background's null cone for defining the equal times in equal-time commutation relations.) Causality difficulties for higher spin fields are not unprecedented.  Long ago Velo and Zwanziger, studying the  spin $\frac{3}{2}$ field, found that ``[t]he main lesson to be drawn from our analysis is that special relativity is not automatically satisfied by writing equations that transform covariantly.  In addition, the solutions must not propagate faster than light.'' \cite{VeloZwanziger}  One proposed solution to the causality problem for massive gravity is to stipulate that solutions with the wrong relationship between the two null cones are \emph{ipso facto} unphysical \cite{LogunovBook}.  This strategy is reasonable, as long as obviously physical solutions satisfy the causality criterion.  

However, for massive gravity (including massive spin 0 here), the static gravitational field from sources decays exponentially in Yukawa form, whereas radiation decays only as $\frac{1}{r}.$ Far from sources, gravitational radiation will resemble a linearized gravity plane wave, which violates $\eta$-causality \cite{NullCones1}, and the static field from the source will have decayed away exponentially and so be negligible.  In response  Chugreev has invoked the infinite distribution of matter in some standard cosmological models \cite{ChugreevCause}.  While this move might perhaps save causality in the actual world, it hardly rescues other presumably physical solutions, such as  finite-range analogs of the Schwarzschild or Kerr-Newman solutions, or solutions with a radiating localized source.  Perhaps the whole universe is in fact filled with matter, but that is certainly not a necessary truth; it might well have turned out that matter was bounded.  

If rejecting bad solutions by hand leaves too few, the alternative \cite{NullCones1} would seem to involve the introduction of gauge freedom so that the relationship between the two null cones is not rigidly fixed by the physical state of the world. Several ways to convert massive  GR to  a gauge theory might be considered.  One way is by parametrization, that is, the introduction of clock fields \cite{SchmelzerMass,Arkani}.  In effect one turns un-variational  Cartesian coordinates into variational clock fields by replacing $\eta_{\mu\nu}$ in field equations (but not in the boundary conditions or notion of causality) in terms of clock fields $X^{A}$ by  $\eta_{\mu\nu} \rightarrow X^{A},_{\mu} \eta_{AB}  X^{B},_{\nu},$  where $\eta_{AB}=diag(-1,1,1,1).$  For the $n=1,$ $l=1$ FMS theory,  $\mathcal{L}$ then  looks like GR plus 4 minimally coupled scalars (one with the wrong sign), but the boundary conditions are  different.  After a change of variables, the resulting gauged massive GR resembles Proca electromagnetism gauged with Stueckelberg's trick. For other massive gravities, the outcome will be more complicated.   Presumably other conversion technologies, such as BFT or Gauge Unfixing, could also be used.

 The gauge freedom could be used to \emph{make} the correct null cone relationship happen in suitably many cases \cite{NullCones1}. The naive gauge transformations act on the two metrics as   $g_{\mu \nu } \rightarrow e^{\pounds _{\xi }}g_{\mu \nu }, \eta _{\mu \nu}\rightarrow \eta _{\mu \nu }. $   One can then redefine gauge transformations so that only $\eta$-causality respecting states are connected by gauge transformations.  Then $\eta$-causality holds by construction.  Thus those transformations that connect a state violating $\eta$-causality and one satisfying it are not truly gauge transformations.  In other words,    $g_{\mu\nu}$ must respect $\eta$-causality, so not every vector field $\xi ^{\mu }$ generates a gauge transformation for every $g_{\mu \nu }$ and $\eta_{\mu \nu }$.  Gauge transformations can be identified not with a generating vector field only, but with an ordered triple involving the metrics:  $ (e^{\pounds _{\xi }},\eta _{\mu \nu },g_{\mu \nu }),$   where $g_{\mu \nu }$  and $e^{\pounds _{\xi }}g_{\mu \nu }$ satisfy $\eta$-causality.   Two gauge transformations $(e^{\pounds_{\psi} }, \eta_{2}, g_{2})$, $(e^{\pounds_{\xi} }, \eta_{1}, g_{1})$ can be composed to give a new gauge transformation
$ (e^{\pounds_{\psi} }, \eta_{2}, g_{2}) \circ (e^{\pounds_{\xi} }, \eta_{1}, g_{1})  =  ( e^{\pounds_{\psi} } e^{\pounds_{\xi} },  \eta_{1}, g_{1})  
$ if and only if the flat metrics match  ($\eta_{2} = \eta_{1}$) and consecutive curved metrics match ($g_{2} = e^{\pounds_{\xi} } g_{1}$.)    Not every pair of gauge transformations can be composed, so
gauge transformations form not a group, but a groupoid (in the sense of Brandt). According to  Ramsay,  ``[a] groupoid is, roughly speaking, a set with a not everywhere defined binary operation, which would be a group if the operation were defined everywhere.'' \cite{Ramsay} The  formal definition  is expressed by  Renault \cite{Renault} as follows: a   groupoid is a  set $G$  with  a product map $(x,y)\rightarrow
xy:G^{2}\rightarrow G$, where $G^{2}  \subset G\times G$ is the set of composable ordered pairs, and the inverse map $x\rightarrow
x^{-1}:G\rightarrow G$ is such that:
\begin{enumerate}
\item  $(x^{-1})^{-1}=x$,
\item  if $(x,y)$ and $(y,z)$ are elements of $G^{2}$, then $(xy,z)$ and $(x,yz)$ are elements of $G^{2}$ and $(xy)z=x(yz),$   
\item  $(x^{-1},x)\in G^{2}$, and if $(x,y)\in G^{2}$, then $x^{-1}(xy)=y$,
\item  $(x,x^{-1})\in G^{2}$, and if $(z,x)\in G^{2}$, then $(zx)x^{-1}=z$. 
\end{enumerate}  Thus there is no single identity element for all elements, but an element times its inverse gives one of many little identity elements.  There are as many little identity elements as one needs; they behave in the expected fashion.  

 By construction, gauge transformations as defined here  preserve the \emph{qualitative} relationship between null cones ($g$ inside $\eta$), while quantitative part is gauge fluff.  (Analogously,  Motter  restricted coordinate transformations in principled way, so ``relativistic chaos is coordinate invariant'':  the Lyapunov exponent's \emph{sign} is invariant, though its value is not  \cite{GRchaos}.)   This relationship is based described using a generalized eigenvalue formalism based on a Segr\'{e} classification of $g_{\mu\nu}$  with respect to $\eta_{\mu\nu}$ \cite{NullCones1}.  In cases satisfying $\eta$-causality (and not arbitrarily close to violating it), there exist four linearly independent eigenvectors with real positive eigenvalues, one of which eigenvectors is timelike and has the smallest eigenvalue. (For generalized eigenvalue problems with an indefinite metric, or alternatively for an eigenvalue problem with an asymmetric matrix, the usual theorems do not apply.) Thus gauge transformations can rotate and boost the eigenvectors in various ways and change the eigenvalues, but cannot change the number or type of eigenvectors and eigenvalues. 
Thus it appears that massive gravity need not be acausal, but achieving causality is nontrivial.


\section*{References}

%


\begin{thebibliography}{10}

\bibitem{EinsteinPapers4}
 Klein M J,  Kox A J, Renn J, and Schulmann R 1995
\newblock {\em The Collected Papers of Albert Einstein, Volume 4, The Swiss
  Years: Writings, 1912-1914} (Princeton: The Hebrew University of Jerusalem and Princeton University)
 
\bibitem{EinsteinPapers8}
 Schulmann R,  Kox A  J,  Janssen M, and Illy J 1998
\newblock {\em The Collected Papers of Albert Einstein, Volume 8, The Berlin
  Years: Correspondence, 1914-1918} (Princeton:  The Hebrew University of Jerusalem and Princeton University)

\bibitem{Janssen}
 Janssen M 2005
\newblock Of pots and holes: {E}instein's bumpy road to general relativity
\newblock {\em Annalen der Physik} {\bf 14:Supplement}  58

\bibitem{PrincipleRelativity}
 Lorentz H A, Einstein A et al 1952
\newblock In {\em The Principle of Relativity} (New York: Dover)  pp 111-173

\bibitem{NullCones}
 Pitts J B and  Schieve W C 2002
\newblock Null cones in {L}orentz-covariant general relativity
\newblock {\em Preprint} gr-qc/0110004

\bibitem{Kraichnan}
 Kraichnan R H 1955
\newblock Special-relativistic derivation of generally covariant gravitation
  theory
\newblock {\em Physical Review} {\bf 98} 1118

\bibitem{Gupta}
 Gupta S N 1954
\newblock Gravitation and electromagnetism
\newblock {\em Physical Review} {\bf 96 } 1683

\bibitem{Feynman}
Feynman R P et al 1995 
\newblock {\em Feynman Lectures on Gravitation}
\newblock (Reading, Mass.: Addison-Wesley) 

\bibitem{Deser}
 Deser S 1970
\newblock Self-interaction and gauge invariance
\newblock {\em General Relativity and Gravitation} {\bf 1} 9
(\emph{Preprint}  gr-qc/0411023)

\bibitem{OP}
 Ogievetsky V I and  Polubarinov I V 1965
\newblock Interacting field of spin 2 and the {E}instein equations
\newblock {\em Annals of Physics} {\bf 35} 167

\bibitem{FMS}
 Freund P G O,  Maheshwari A, and  Schonberg E 1969
\newblock Finite-range gravitation
\newblock {\em Astrophysical Journal} {\bf 157} 857

\bibitem{PittsMassive}
 Pitts  J B and  Schieve W C 2005
\newblock Universally coupled massive gravity, {I}
\newblock {\em Preprint} gr-qc/0503051

\bibitem{Visser}
 Visser M 1998
\newblock Mass for the graviton
\newblock {\em General Relativity and Gravitation} {\bf  30} 1717
(\emph{Preprint}  gr-qc/9705051v2)

\bibitem{GrishchukMass}
 Babak S V and  Grishchuk L P 2003
\newblock Finite-range gravity and its role in gravitational waves, black holes
  and cosmology
\newblock {\em International Journal of Modern Physics D} {\bf  12} 1905
(\emph{Preprint}  gr-qc/0209006)

\bibitem{vDVmass2}
 van Dam H and  Veltman M 1972
\newblock On the mass of the graviton
\newblock {\em General Relativity and Gravitation} {\bf 3} 215

\bibitem{Zakharov}
 Zakharov V I 1970
\newblock Linearized gravitation theory and the graviton mass
\newblock {\em Journal of Experimental and Theoretical Physics Letters} {\bf  12} 312

\bibitem{DeserMass}
 Boulware D G and  Deser S 1972
\newblock Can gravitation have a finite range?
\newblock {\em Physical Review D} {\bf  6} 3368

\bibitem{Vainshtein2}
 Deffayet C,  Dvali G,  Gabadadze G, and  Vainshtein A I  2002
\newblock Nonperturbative continuity in graviton mass versus perturbative
  discontinuity
\newblock {\em Physical Review D} {\bf 65} 044026
(\emph{Preprint}  hep-th/0106001)

\bibitem{NullCones1}
 Pitts J B and  Schieve W C 2004
\newblock Null cones and {E}instein's equations in {M}inkowski spacetime
\newblock {\em Foundations of Physics} {\bf 34} 211
(\emph{Preprint} gr-qc/0406102)

\bibitem{LogunovBook}
 Logunov A A 1998
\newblock {\em The Relativistic Theory of Gravitation}
\newblock (Commack, NY: Nova Science)

\bibitem{Marzban}
Marzban C, Whiting B F and van Dam H 1989
\newblock Hamiltonian reduction for massive fields coupled to sources
\newblock {\em Journal of Mathematical Physics} {\bf  30} 1877

\bibitem{Sundermeyer}
 Sundermeyer K 1982
\newblock {\em Constrained Dynamics}
\newblock (Berlin: Springer)

\bibitem{Penrose}
 Penrose R 1980
\newblock   {\em Essays in General Relativity---A
  Festschrift for Abraham Taub} F.~J. Tipler ed (New York: Academic)
\newblock On {S}chwarzschild causality -- a problem for ``{L}orentz covariant''
  general relativity

\bibitem{Zel2}
 Zel'dovich Ya B and  Grishchuk L P 1988
\newblock The general theory of relativity is correct!
\newblock {\em Soviet Physics Uspekhi} {\bf 31} 666

\bibitem{VeloZwanziger}
 Velo G and Zwanziger D 1969
\newblock Propagation and quantization of {R}arita-{S}chwinger waves in an
  external electromagnetic field
\newblock {\em Physical Review} {\bf 186} 1337

\bibitem{ChugreevCause}
 Chugreev Yu V 2004
\newblock Do gravitational waves violate a causality principle?
\newblock {\em Theoretical and Mathematical Physics} {\bf  138} 293

\bibitem{SchmelzerMass}
Schmelzer I 1998
\newblock General ether theorie [\emph{sic}] and graviton mass
\emph{Preprint}  gr-qc/9811073

\bibitem{Arkani}
 Arkani-Hamed N,  Georgi H, and  Schwartz M D 2003
\newblock Effective field theory for massive gravitons and gravity in theory
  space
\newblock {\em Annals of Physics} {\bf 305} 96
(\emph{Preprint}  hep-th/0210184)

\bibitem{Ramsay}
 Ramsay A 1971
\newblock Virtual groups and group actions
\newblock {\em Advances in Mathematics} {\bf 6} 253

\bibitem{Renault}
 Renault J   1980
\newblock {\em A Groupoid Approach to $C^{*}$-algebras}
\newblock {Berlin: Springer}

\bibitem{GRchaos}
 Motter A E 2003
\newblock Relativistic chaos is coordinate invariant
\newblock {\em Physical Review Letters} {\bf 91} 231101
(\emph{Preprint} gr-qc/0305020)

\end{thebibliography}

\end{document}